\documentclass[aip,jcp,amsmath,amssymb,amsfonts,reprint,longbibliography,draft]{revtex4-1}

\usepackage{graphicx}
\usepackage{color}
\usepackage[normalem]{ulem}

\DeclareMathAlphabet{\mathbf}{OT1}{cmr}{bx}{it}
\DeclareMathAlphabet{\mathssb}{OT1}{cmss}{bx}{n}
\DeclareMathAlphabet{\mathssn}{OT1}{cmss}{m}{n}
\DeclareMathAlphabet{\mathub}{OT1}{cmr}{b}{n}

\newcommand{\s}{\mathbf s}

\newcommand{\f}{\mathbf f}

\newcommand{\p}{\mathbf p}
\newcommand{\q}{\mathbf q}

\renewcommand{\P}{{\mathbf P}}
\newcommand{\T}{{\mathbf T}}
\renewcommand{\r}{\mathbf r}

\newcommand{\R}{{\mathbb R}}

\renewcommand{\v}{\mathbf v}

\newcommand{\F}{\mathbf F}
\newcommand{\G}{\mathbf G}

\newcommand{\I}{\mathbf I}

\newcommand{\1}{\mathbf{1}}
\newcommand{\x}{\mathbf{x}}

\newcommand{\dvg}[1]{\operatorname{\mathrm{div}_\mathit{#1}}}

\def\Real{{\mathord{\rm{I \kern-.22em R}}}}

\newcommand{\de}{\partial}
\newcommand{\tsp}{\mathsf{T}\mskip-3mu}

\begin{document}

\title{Continuum balances from extended Hamiltonian dynamics}
\author{Giulio G.\ Giusteri}
\email{giulio.giusteri@oist.jp}
\affiliation{Mathematical Soft Matter Unit, Okinawa Institute of Science and Technology Graduate University, 1919-1 Tancha, Onna, Okinawa, 904-0495, Japan}
\author{Paolo Podio-Guidugli}
\email{p.podioguidugli@gmail.com}
\affiliation{Accademia Nazionale dei Lincei, Palazzo Corsini, \mbox{Via della Lungara 10, 00165 Roma, Italy}}
\affiliation{Department of Mathematics, Universit\`a di Roma TorVergata, \mbox{Via della Ricerca Scientifica 1, 00133 Roma, Italy}}
\author{Eliot Fried}
\email{eliot.fried@oist.jp}
\affiliation{Mathematical Soft Matter Unit, Okinawa Institute of Science and Technology Graduate University, 1919-1 Tancha, Onna, Okinawa, 904-0495, Japan}

\date{\today}

\begin{abstract}
The classical procedure devised by Irving \& Kirkwood in 1950 and completed slightly later by Noll produces counterparts of the basic balance laws of standard continuum mechanics starting from an ordinary Hamiltonian description of the dynamics of a system of material points. Post-1980 molecular dynamics simulations of the time evolution of such systems use extended Hamiltonians such as those introduced by Andersen, Nos\'e, and Parrinello and Rahman. The additional terms present in these extensions affect the statistical properties of the system so as to capture certain target phenomenologies that would otherwise be beyond reach. We here propose a physically consistent application of the Irving--Kirkwood--Noll procedure to extended Hamiltonian systems of material points. Our procedure produces balance equations at the continuum level featuring non-standard terms, because the presence of auxiliary degrees of freedom gives rise to additional fluxes and sources that influence the thermodynamic and transport properties of the continuum model. Being aware of the additional contributions may prove crucial when designing multiscale computational schemes in which information is exchanged between the atomistic and continuum levels.
\end{abstract}

\maketitle

\section{Introduction}

Whereas classical continuum mechanics rests on the description of the deterministic evolution of macroscopic quantities, classical statistical mechanics studies probabilistic features of microscopic observables.
Continuum mechanics provides phenomenological descriptions typically aimed
at determining over a time interval and a matter-comprised
space region the values taken by a list of fields, such as 
mass density, motion, stress, and temperature. This is done by solving
initial/boundary-value problems governed by sets of partial-differential
evolution equations embodying such fundamental
laws as the balances of mass, linear momentum, and total
energy, specialized for a constitutive class of choice. Statistical mechanics provides expressions for equations of state and forces based on a chosen potential of internal interactions in combination with the ensemble properties. Their different natures notwithstanding, these edifices can supply consistent and complementary information when applied to one and the same physical system with the objective of building a multiscale method. When the statistical information retrieved at the atomistic level is used to obtain deterministic predictions at the continuum level, or vice versa, there are two main consistency issues to settle: one concerning macroscopic constitutive responses to deformations and microscopic interatomic potentials; the other concerning the evolution equations in the two frameworks. In this paper, we deal with the latter issue.

To close the atomistic-to-continuum scale gap to the extent of associating a set of deterministic balances with a probabilistic account of the Newtonian evolution of a system of material points, we take the bottom-up path pioneered by Irving \& Kirkwood \cite{IK} in a paper that appeared in 1950. Their path was completed a few years later by Noll,\cite{N55,*NT10} who derived closed-form integral expressions relating microscopic observables to the continuum mechanical notions of stress and heat flux.

The Irving--Kirkwood--Noll (IKN) procedure was derived for the classical Hamiltonian dynamics of a collection of material points. This represents a significant limitation, because the statistical information that can be retrieved via Molecular Dynamics (MD) from a classical Hamiltonian system refers only to the microcanonical ensemble (NVE), in which the number of particles, the volume, and the energy are conserved.

An innovative and effective technique for accessing different statistical ensembles was introduced by Andersen\cite{HCA} in 1980. Andersen's idea was to view the MD computational cell as a carrier of mesoscopic information, conveyed by an extended Hamiltonian, in which the introduction of an additional dynamical variable---the volume of the cell in his case---allows the system to explore the phase space along trajectories for which the value of the extended Hamiltonian, but not the physical energy, is conserved. As a consequence, while the extended system retains statistics pertinent to the extended microcanonical ensemble, the statistical ensemble induced on the physical degrees of freedom is modified to obtain the conservation of two observables---the pressure applied to the cell and the related enthalpy. Prompted by Andersen's paper, Parrinello \& Rahman\cite{PR1,*PR2} let not only the volume but also the shape of the computational cells change. They showed  that the availability of shape-related degrees of freedom allows for simulating stress-induced phase transitions in solids; to paraphrase the words of Nos\'e,\cite{Nos91} their papers  revolutionized the approach to investigating structural phase transitions. A few years later, Nos\'e,\cite{Nos84b,*Nos84} inspired once again by Andersen's paper, conceived the idea of an extended Hamiltonian befitting MD simulations consistent with a constant-temperature canonical ensemble (NVT), in the sense that ``the canonical  distribution is realized in a physical system''.\cite{Nos91} Simulations of extended Hamiltonian dynamics were also considered within the more general class of non-Hamiltonian MD in important articles on their statistical properties.\cite{TucMun99,*TucLiu01} A development that, to our knowledge, went unnoticed in the literature is that Noll,\cite{N55} whose treatment allows for the presence of non-conservative external forces, actually considered the extension to a special class of non-Hamiltonian evolutions in as early as 1955. 

In the present paper, we couple the IKN procedure---in itself a microscopic-to-macroscopic scale-bridging procedure---with two extended Hamiltonians.
%
%
The first models the Nos\'e--Hoover (NH) thermostat.\cite{Nos84b,*Nos84,Nos91,Hoo85,*Hoo07}
The second is a Hamiltonian of Andersen--Parrinello--Rahman (APR) type, which  includes a complete accounting of the kinetic energy. The extended Lagrangian used by Parrinello \& Rahman,\cite{PR1,*PR2} which featured a simplified accounting of the kinetic energy, was a generalization of that previously devised by Andersen with the objective of achieving an isobaric ensemble,\cite{HCA} whence our use of the acronym APR.

As is well-known, the IKN procedure tells us what statistical observables to associate with the continuum mechanical notions of mass, linear momentum, and energy density. On evolving these observables \emph{\`a la} Liouville with respect to a chosen Hamiltonian, three consequences of the Newtonian motion of a system of material points are derived. These can be associated with three balance laws basic to continuum mechanics, namely mass conservation, balance of linear momentum, and balance of energy. The association involves convenient identifications of other statistical observables with the continuum mechanical notions of stress and heat flux. In recent years, there has been a renewed interest in these matters; with no pretense to exhaustivity, we mention the works of Murdoch\cite{Mur10,*Mur12} (who proposes an identification procedure alternative to that used in the IKN procedure along lines anticipated by Hardy\cite{Har82}), Admal \& Tadmor,\cite{AdmTad10,*AdmTad11,*AdmTad16} Seguin \& Fried,\cite{SegFri12,*SegFri13} and others.\cite{WajAlt95,DenRob04,*DenRob06,Gol10,HumTok11,LehSea11,YangWuLi12,CapMar14,DavSte14a,*DavSte14b,*DavSte15}

We remark that in the case of microstructured continua a microscopic definition of angular momentum is often necessary and would significantly affect the continuum balance of the density of angular momentum, as recently discussed by Seguin \& Fried\cite{SegFri12,*SegFri13} and Davydov \& Steinmann.\cite{DavSte14a,*DavSte14b,*DavSte15} Nevertheless, in the present treatment, we restrict attention to continua associated with systems of material points that do not possess rotational degrees of freedom. In such a context, the continuum balance of angular momentum is equivalent to the symmetry of the Cauchy stress tensor, a property guaranteed under fairly general assumptions concerning the nature of the internal interactions.

Our goal is to implement the IKN procedure for both the NH and the APR Hamiltonians. Our main expectation is to provide an indication as to which continuum equations (and of what kind) should be coupled to  MD simulations when building consistent multiscale schemes. Our interest in the NH thermostat is motivated by the assumption of isothermal evolution, posited for many continuum mechanical models. On the other hand, the APR approach has the peculiarity of introducing a dynamic generalization of the Cauchy--Born rule,\cite{ppg1,ppg2} which makes it directly relevant to certain recently proposed multiscale computational schemes.\cite{DavPelSte14,Ulz15,LiTong15,*TongLi15} 
A discussion of how thermostatting techniques can affect the evaluation of transport coefficients in linear response theory (which are related to the balance of linear momentum) is presented by Evans \& Morriss.\cite{EvaMor08} Our goal, however, is to obtain a full set of continuum balances featuring a more accurate account of the macroscopic effects induced by the extended Hamiltonian dynamics.

When setting up the IKN procedure for an extended Hamiltonian system, that is to say, a particle system whose Hamiltonian $H$ is nonstandard, two key observations are that (i) certain microscopic observables must be defined by combining ordinary physical variables with other possibly ``unphysical'' variables and that (ii) all macroscopic observables, whatever their microscopic antecedents, consist of ensemble averages weighted with respect to probability densities defined over the extended phase space. This in itself guarantees a basic statistical-continuum consistency in the derived balances of macroscopic observables. However, we find that an extended Hamiltonian evolution always entails the presence of new terms in the balance of energy. This signals that the choice of a peculiar probabilistic dynamics at the microscopic scale modifies the thermodynamic and transport properties of the system at the macroscopic scale. Awareness of this denouement is crucial when devising computational schemes in which information is exchanged between the macroscopic and microscopic levels.


Our paper is organized as follows. In Section~\ref{sec:Liouville}, we recall a few well-known notions regarding Liouville's evolution of probability densities and observables for Hamiltonian particle systems; we also specify the changes in a format necessary to deal with extended Hamiltonian systems. In Section~\ref{sec:extended}, we discuss the application to extended Hamiltonian systems of the IKN procedure for obtaining macroscopic balance equations; in particular, we analyze in detail the cases of the NH thermostat and of APR dynamics. The main results of our study, as well as their relevance to multiscale computational schemes, are finally summarized in Section~\ref{sec:conclusions}.

\section{Liouville's evolution of observables}\label{sec:Liouville}

Consider a system defined by a Hamiltonian $H$ depending on Lagrangian coordinates $q$ and conjugate momenta $p$, with $z=(q,p)$ denoting a generic point in the phase space $\mathcal Z$. The evolution of the system is given by Hamilton's equations
\begin{equation}\label{eq:Hamilton}
\dot{\zeta}(t)=\{z,H\}\vert_{z=\zeta(t)},
\end{equation}
where $\{\cdot,\cdot\}$ denotes the Poisson bracket defined, for any pair $A$ and $B$ of functions of $(z,t)=(q,p,t)$, as
\begin{equation}
\{A,B\}:=\frac{\de A}{\de q}\cdot\frac{\de B}{\de p}-\frac{\de A}{\de p}\cdot\frac{\de B}{\de q}.
\end{equation}

\subsection{Liouville's theorem and Liouville's equation}\label{LTh}

Given a Hamiltonian system, it is readily checked that
\begin{equation}\label{eq:incompressible}
\dvg{z}\{z,H\}=0,
\end{equation}
which is a statement of Liouville's theorem. A geometrical statement of Liouville's theorem based on \eqref{eq:Hamilton} and \eqref{eq:incompressible} is that any Hamiltonian evolution of a region in the phase space $\mathcal Z$ is locally volume-preserving.

The incompressibility condition expressed by \eqref{eq:incompressible} has another important consequence: any probability density function $\tilde\rho$ defined on $\mathcal Z$ at the initial time $t_0$ is convected along Hamiltonian trajectories in the sense that
\begin{equation}\label{eq:LiouvilleM}
\dot{\tilde{\rho}}(\zeta(t),t)=0
\end{equation}
whenever $\zeta(t)$ is a solution of \eqref{eq:Hamilton}.
This can be equivalently expressed by Liouville's equation
\begin{equation}\label{eq:Liouville}
\frac{\de}{\de t}\tilde\rho(z,t)-\{H(z,t),\tilde{\rho}(z,t)\}=0,
\end{equation}
where 
\begin{equation}\label{liuop}
\ell[\cdot]:=-\{H(z,t),\cdot\}
\end{equation}
is the Liouville operator for a system of Hamiltonian $H$.
Liouville's equation evolves a given initial value $\tilde\rho(z,t_0)=\tilde\rho_0(z)$ of the probability measure at any given point of $\mathcal Z$ into its current value $\tilde\rho(z,t)$ at that same point. The kernel of the Liouville operator $\ell$ consists of all the stationary probability measures, that is, those which are independent of time $t$. Notice that the discussion of the present section applies also to time-dependent Hamiltonians, even though, in the applications discussed subsequently, we only encounter time-independent Hamiltonians.

\subsection{Extended form of Liouville's equation}

If we consider a decomposition $\mathcal Z=\mathcal Z_\mathrm{s}\times\mathcal Z_\mathrm e$ of the phase space $\mathcal Z$, where $\mathcal Z_\mathrm{s}$ is associated with a subset of Hamiltonian coordinates, denoted by $(q_\mathrm{s},p_\mathrm{s})$, and $\mathcal Z_\mathrm e$ is associated with additional degrees of freedom that extend the set of coordinates,
we can then define the restricted Poisson bracket for any pair $A$ and $B$ of scalar functions of $(z,t)$ by
\begin{equation}
\{A,B\}_{\mathcal Z_\mathrm{s}}:=\frac{\de A}{\de q_\mathrm{s}}\cdot\frac{\de B}{\de p_\mathrm{s}}-\frac{\de A}{\de p_\mathrm{s}}\cdot\frac{\de B}{\de q_\mathrm{s}};
\end{equation}
the complementary bracket $\{A,B\}_{\mathcal Z_\mathrm{e}}$ has an analogous expression. 
With such definitions, we obtain an additive decomposition of the Poisson bracket:
\begin{equation}\label{eq:decomp}
\{A,B\}=\{A,B\}_{\mathcal Z_\mathrm{s}}+\{A,B\}_{\mathcal Z_\mathrm{e}}.
\end{equation}
Moreover, we can exploit \eqref{eq:decomp} to express Liouville's evolution of probability densities for an extended Hamiltonian system as
\begin{equation}\label{eq:LiouvilleE}
\frac{\de\tilde\rho}{\de t}=\{H,\tilde{\rho}\}_{\mathcal Z_\mathrm{s}}+\{H,\tilde{\rho}\}_{\mathcal Z_\mathrm{e}}.
\end{equation}

\subsection{Extended evolution of observables}

A {microscopic observable} is a mapping $\{z\mapsto \hat{o}(z)\}$ defined over the phase space $\mathcal Z$. At this stage, the tensorial order of $\hat o$ can be left arbitrary. The corresponding {macroscopic observable} is the {expected value} of $\hat{o}$ at time $t$ with respect to the chosen probability density $\tilde\rho(z,t)$, namely 
\begin{equation}\label{obs}
\langle o\rangle(t):=\int_{\mathcal Z}\tilde\rho(z,t)\hat o(z)\,dz.
\end{equation}

An important application of Liouville's equation is to compute the time derivative of the expected value $\langle o\rangle$ of a microscopic observable $\hat{o}$, which is given by
\begin{equation}\label{eq:obs-evolution}
\frac{\de\langle o\rangle}{\de t}
=\int_{\mathcal Z} \{H,\tilde{\rho}\}\hat o\,dz.
\end{equation}
For extended Hamiltonian systems, substituting the additive decomposition \eqref{eq:decomp} in \eqref{eq:obs-evolution} gives
\begin{equation}\label{eq:obs-evolutionE}
\frac{\de\langle o\rangle}{\de t}=\int_{\mathcal Z} \{H,\tilde{\rho}\}_{\mathcal Z_\mathrm{s}}\hat o\,dz+\int_{\mathcal Z} \{H,\tilde{\rho}\}_{\mathcal Z_\mathrm{e}}\hat o\,dz.
\end{equation}

Irving \& Kirkwood\cite{IK} (in a footnote on their p.~822) commented on the possible presence of additional degrees of freedom, leading to \eqref{eq:obs-evolutionE}. However, they considered only observables that are independent of the additional degrees of freedom. At variance with them, we will consider observables that depend also on such coordinates, entailing non-vanishing contributions from both terms in the right-hand side of \eqref{eq:obs-evolutionE}.

\section{The IKN procedure for extended Hamiltonian systems}\label{sec:extended}\label{III}

We begin by briefly recapping the classical IKN procedure, which 
is applied to a collection of $N$ material points moving in  three-dimensional  space. The Hamiltonian coordinates are
\begin{equation}\label{IKobs1}
q=(\r_1,\ldots,\r_N)
\quad\text{and}\quad
p=(m_1\dot\r_1,\ldots,m_N\dot\r_N),
\end{equation}
where $\r_k$ is the current position vector, with respect to a fixed origin, of the $k\textrm{-th}$ material point, the mass of which is $m_k$.
The fundamental presumption is that all macroscopic properties of matter are deducible from a microscopic discrete picture; the idea is to exploit the similarity in format of continuum balance equations and statistical Liouville flows. 

The procedure consists of three steps: (i) The densities of mass, linear momentum, kinetic energy, and internal energy are associated with the ensemble averages of certain microscopic observables, the common form of which is
\begin{equation}\label{IKobs}
\hat o(z,\r)=\sum_k\hat o_k(z)\,\delta(\r_k-\r),
\end{equation}
where  $\r$ is the position vector of the typical space point. (ii) Each of the above ensemble averages  is evolved, in the order given, \emph{\`a la} Liouville. (iii) The so-obtained Liouville flows are identified term-by-term with their continuum mechanical counterparts. An all important feature of the procedure is that the representation \eqref{IKobs} of each microscopic observable---a finite sum of Dirac distributions, each supported at one of the material points constituting the system---implies that the associated macroscopic observable
\begin{align}
\langle o\rangle(\r,t)&=\Big\langle \,\sum_k\hat o_k(z)\delta(\r_k-\r)\,\Big\rangle\notag\\
&
=\int_{\mathcal Z} \hat o(z,\r)\,\tilde\rho(z,t)\,dz\label{micmac}
\end{align}
is a time-dependent spatial field.

When applying the IKN procedure to an extended Hamiltonian system, it is important that the basic macroscopic observables maintain their physical meaning. This poses restrictions on the definitions of the relative microscopic observables, where possibly `unphysical' additional Hamiltonian coordinates inevitably enter. Moreover, the additional Hamiltonian coordinates affect the structure of both the phase space and the probability measure entering ensemble averages. Altogether, it can happen that certain terms of the continuum balances produced by the IKN procedure turn out to have forms different from the usual ones. In the following, we exemplify these developments for two paradigmatic extended Hamiltonian systems.

\subsection{Application to the NH thermostat}

The first system to which we apply the IKN procedure is the extended Hamiltonian system introduced by Nos\'e,\cite{Nos84b,*Nos84,Nos91} which forms the basis for the thermostatting strategy proposed by Hoover.\cite{Hoo85,*Hoo07} We shall see, a physically sound definition of  observables produces macroscopic balances that are consistent with classical results. Nevertheless, additional source terms that stem from the extended Hamiltonian are present both in the mass balance and in the energy balance. At the continuum level, those additional terms can alter the spatial and temporal distributions of the fields. At the microscopic level, it is, however, the ensemble that changes.

\subsubsection{The extended Hamiltonian}
Nos\'e's extended Hamiltonian can be written as
\begin{multline}\label{eq:Hnh}
H_\mathrm{NH}(q^\prime,p^\prime;Q,T):=
\sum_k\frac{|\p_k|^2}{2m_ks^2}+\frac{p_s^2}{2Q}+V(q)\\
+(3N+1)k_\mathrm{B}T(\ln s-1),
\end{multline}
the underlying coordinates being
\begin{equation}
q^\prime=(q,s)\quad\textrm{and}\quad p^\prime=(\p_1,\ldots,\p_N,p_s).
\end{equation}
The string $q^\prime$ is the standard string of Lagrangian coordinates in  $\eqref{IKobs1}_1$, augmented by a dimensionless scalar coordinate $s$; $p_s$, the momentum conjugate to $s$, augments  a string  of virtual momenta $\p_k,\;k\in\{1,\ldots,N\}$. 
For the physical velocity $\boldsymbol{v}_k$ of the $k$-th material point, the relationship between the virtual momentum $\p_k$ and the physical momentum $m_k\v_k$ is mediated by $s$ as follows:
\begin{equation}
\frac{\p_k}{s}=m_k\v_k.
\end{equation}
The role assigned by Nos\'e to $s$ is to scale the physical time $t$ in terms of the virtual time $\tau$ involved in the Hamiltonian evolution associated with $H_\mathrm{NH}$ according to: 
\begin{equation}\label{eq:realtime}
t=\int_0^\tau\frac{d\alpha}{s(\alpha)}.
\end{equation}
This interpretation of $s$, which depends on the virtual time, is consistent with Nos\'e's intention to achieve temperature control by controlling particle velocities, granted the understanding that the temperature of a system is related to the average of its kinetic energy.\cite{Nos91} For this formulation to make sense,  $s$ must be positive. Consistent with this observation, we stipulate that the probability density function $\tilde\rho(q^\prime,p^\prime,\tau)$ must vanish identically for $s\leq 0$.
 
The expression of $H_\mathrm{NH}$ in \eqref{eq:Hnh} features two parameters: $Q$, which enters the new kinetic term; and $T$, which enters the new potential energy, where $N$ is the number of material points and $k_\mathrm{B}$ is Boltzmann's constant. Specifically, $Q$ modulates the new kinetic term proportional to $p_s^2$ and plays the role of an effective thermal inertia, associated with the characteristic frequency of thermal fluctuations, and $T$ is the target temperature. The role of the `entropic' potential $(3N+1)k_\mathrm{B}T(\ln s-1)$ is to induce the constant-temperature ensemble characterizing the equilibrium properties of the system.

Finally, in \eqref{eq:Hnh} the potential energy $V$ has the standard form $V=V^\mathrm{i}+V^\mathrm{e}$, where the internal contribution
\begin{equation}\label{eq:Vi}
V^\mathrm{i}(q):=\frac{1}{2}\sum_{j\neq k}V_{jk}(|\r_j-\r_k|)
\end{equation}
accounts for pairwise interactions between material points and 
\begin{equation}\label{eq:Ve}
V^\mathrm{e}(q):=\sum_k V^\mathrm{e}_k(\r_k),
\end{equation}
is the unary potential of external forces.

\subsubsection{Definition of the observables}

We now identify the statistical counterparts of the fields entering the basic continuum balances as macroscopic observables under the form of ensemble averages of suitable physically relevant microscopic observables.  {We do so by applying the general IKN prescription recalled in the introductory paragraph of this section. By construction, {such ensemble averages} depend on the virtual time $\tau$; in what follows, the inversion of relation \eqref{eq:realtime} is implied. Specifically, we define the mass density
\begin{equation}\label{eq:rhonh}
\rho(\r,t)\equiv\Big\langle\sum_k m_k\delta(\r_k-\r)\Big\rangle,
\end{equation}
the linear momentum density
\begin{align}
\rho\v(\r,t)&\mbox{}\equiv\Big\langle\sum_k \frac{\p_k}{s}\delta(\r_k-\r)\Big\rangle\notag\\
&\mbox{}=\Big\langle\sum_k m_k\v_k\delta(\r_k-\r)\Big\rangle,\label{eq:pnh}
\end{align}
the kinetic energy density
\begin{align}
\epsilon_K(\r,t)&\mbox{}\equiv\Big\langle\sum_k \frac{|\p_k|^2}{2m_ks^2}\delta(\r_k-\r)\Big\rangle\notag\\
&\mbox{}=\Big\langle\sum_k \frac{m_k}{2}|\v_k|^2\delta(\r_k-\r)\Big\rangle,\label{eq:eKnh}
\end{align}
and the potential energy density
\begin{multline}\label{eq:eVnh}
\epsilon_V(\r,t)\equiv\mbox{}\Big\langle \sum_{j\neq k}\frac{1}{2}V_{jk}(|\r_j-\r_k)|)\delta(\r_k-\r)\Big\rangle\\
\mbox{}+\Big\langle\sum_{k}V^\mathrm{e}_k(\r_k)\delta(\r_k-\r)\Big\rangle.
\end{multline}

The presence of additional degrees of freedom and energy terms in $H_\mathrm{NH}$ calls for the definition of further observables.
Since the additional degrees of freedom are not explicitly ascribable to individual material points, it is not immediately obvious whether the corresponding observables should be given the form \eqref{IKobs}, entailing generally nontrivial variations with spatial position. These observables could also be viewed as collective properties of the system and be described by uniform spatial fields. To investigate the implications of those two perspectives, we define both collective observables, which are independent of the space point $\r$, and corresponding distributed observables (indicated with superposed bars) of IKN form.

The collective observables associated with the extra kinetic energy density and the entropic energy density are
\begin{align}
{\epsilon}_{p_s}(t)&\mbox{}\equiv\Big\langle\frac{1}{\omega_\mathrm{ref}}\frac{p_s^2}{2Q}\Big\rangle,\label{eq:eps-til}
\\[4pt]
{\epsilon}_s(t)&\mbox{}\equiv\Big\langle \frac{A}{\omega_\mathrm{ref}}(\ln s-1)\Big\rangle,\label{eq:es-til}
\end{align}
where we have introduced $A:=(3N+1)k_\mathrm{B}T$ and $\omega_\mathrm{ref}$ is used to denote the volume of the computational cell.

The construction of an IKN observable for a collective quantity can always be achieved by multiplying that quantity by the microscopic precursor
\begin{equation}
\sum_k\frac{1}{N}\delta(\r_k-\r)
\end{equation}
of the number density
\begin{equation}
n(\r,t)\equiv\Big\langle\sum_k\frac{1}{N}\delta(\r_k-\r)\Big\rangle.
\end{equation}
Following that prescription, we define the distributed observables
\begin{align}
\bar{\epsilon}_{p_s}(\r,t)&\mbox{}\equiv\Big\langle\frac{p_s^2}{2Q}\sum_k\frac{1}{N}\delta(\r_k-\r)\Big\rangle,\label{eq:kps}\\
\bar{\epsilon}_s(\r,t)&\mbox{}\equiv
\Big\langle A(\ln s-1)\sum_{k}\frac{1}{N}\delta(\r_k-\r)\Big\rangle.\label{eq:esnh}
\end{align}

The foregoing definitions and the corresponding evolution terms represent an important novelty of the continuum theory associated with an extended Hamiltonian system via the IKN procedure. We anticipate that the main difference in choosing collective or distributed observables is that the collective ones, being spatially uniform, are not convected with the motion of material points. For this reason, they produce only uniform source terms in the energy balance, whereas distributed observables produce both convective and diffusive terms in that balance.

\subsubsection{The mass and momentum balances}\label{3}

By applying Liouville's equation \eqref{eq:LiouvilleE}, which in this case takes the form
\begin{multline}
\frac{\de\tilde\rho}{\de \tau}=\sum_k\bigg( \frac{\de H_\mathrm{NH}}{\de\r_k}\cdot\frac{\de\tilde\rho}{\de\p_k}-\frac{\de\tilde\rho}{\de\r_k}\cdot\frac{\de H_\mathrm{NH}}{\de\p_k}\bigg)\\
+\frac{\de H_\mathrm{NH}}{\de s}\frac{\de\tilde\rho}{\de p_s}-\frac{\de\tilde\rho}{\de s}\frac{\de H_\mathrm{NH}}{\de p_s},
\end{multline}
we can deduce the macroscopic balances of the defined observables.

The time derivative of the mass density is
\begin{align}
\frac{\de\rho}{\de t}=\mbox{}&\int_\mathcal Z s\frac{\de\tilde\rho}{\de\tau}\sum_k m_k\delta(\r_k-\r)\,dz\notag\\
=\mbox{}&-\dvg{\r}\int_\mathcal Z\tilde\rho\sum_k \frac{\p_k}{s}\delta(\r_k-\r)\,dz\notag\\
&\mbox{}+\int_\mathcal Z\tilde\rho\frac{p_s}{Q}\sum_k m_k\delta(\r_k-\r)\,dz\notag\\
=\mbox{}&-\dvg{\r}(\rho\v)+\sigma_\rho,\label{divappears}
\end{align}
where the source term is given by
\begin{equation}
\sigma_\rho(\r,t)\equiv\Big\langle\frac{p_s}{Q}\sum_k m_k\delta(\r_k-\r)\Big\rangle.
\end{equation}
As for the linear momentum density, we find that
\begin{align}
\frac{\de(\rho\v)}{\de t}
=\mbox{}&\mbox{}-\int_\mathcal Z\tilde\rho\sum_k \frac{\de V}{\de\r_k}\delta(\r_k-\r)\,dz\notag\\
&\mbox{}-\dvg{\r}\int_\mathcal Z\tilde\rho\sum_k m_k\v_k\otimes\v_k\delta(\r_k-\r)\,dz\notag\\
=\mbox{}&\mbox{}-\dvg{\r}(\rho\v\otimes\v-\T)+\f^\mathrm{e},
\end{align}
where the Cauchy stress tensor $\T$ is the sum of 
the kinetic stress tensor
\begin{equation}\label{eq:TK}
\T_K(\r,t)\equiv-\Big\langle\sum_k m_k(\v_k-\v)\otimes(\v_k-\v)\delta(\r_k-\r)\Big\rangle
\end{equation}
and the stress tensor associated with internal interactions, namely
\begin{equation}\label{eq:TV}
\T_V(\r,t)\equiv\frac{1}{2}\sum_{j\neq k}\int_{\R^3}\frac{\x\otimes\x}{|\x|}V'_{jk}(|\x|)\mathcal J(\x,\r,t)\,d\x,
\end{equation}
with
\begin{equation}\label{eq:TVJ}
\mathcal J(\x,\r,t):=\int_0^1\Big\langle\delta(\r_j-\r-\alpha\x)\delta(\r_k-\r+(1-\alpha)\x)\Big\rangle\,d\alpha,
\end{equation}
and where the external force field is
\begin{equation}\label{eq:fext}
\f^\mathrm{e}(\r,t)\equiv-\Big\langle\sum_k \frac{\de V_k^\mathrm{e}}{\de{\r_k}} \delta(\r_k-\r)\Big\rangle.
\end{equation}

The detailed derivation of formulae \eqref{eq:TV} and \eqref{eq:TVJ} was provided originally by Noll,\cite{N55,*NT10} who more recently generalized and refined his calculations.\cite{N2010}
Note that the very presence of the source term $\sigma_\rho$ in the mass balance, a term that has no classical counterpart, is dictated by the form \eqref{eq:Hnh} of the extended Hamiltonian $H_\mathrm{NH}$, which however does not give rise to specific new terms in the balance of linear momentum.

\subsubsection{Energy balances}
In view of our previous definitions of collective and distributed energy observables, we must balance two energy densities, namely
%
\begin{equation}
\label{eq:energy-density-til}
\epsilon=\epsilon_K+\epsilon_V+\epsilon_{p_s}+\epsilon_s
\end{equation}
and
\begin{equation}
\label{eq:energy-balance-tim}
\bar{\epsilon}=\epsilon_K+\epsilon_V+\bar{\epsilon}_{p_s}+\bar{\epsilon}_s.
\end{equation}
%
In both instances, the time derivatives of $\epsilon_K$ and $\epsilon_V$ produce the counterpart of the continuum mechanical heat flux $\q$, which involves the sum of a kinetic contribution
\begin{equation}\label{eq:qK}
\q_K(\r,t)\equiv\Big\langle\sum_k\frac{m_k}{2}|\v_k-\v|^2(\v_k-\v) \delta(\r_k-\r)\Big\rangle,
\end{equation}
an interaction contribution
\begin{widetext}
\begin{equation}\label{eq:qV}
{\q_V(\r,t)}\equiv-\frac{1}{2}\sum_{j\neq k}\int_{\R^3}\bigg[\frac{\x}{|\x|}V'_{jk}(|\x|)\x\cdot
\int_0^1\Big\langle \frac{1}{2}(\v_j+\v_k)\delta(\r_j-\r-\alpha\x)\delta(\r_k-\r+(1-\alpha)\x)\Big\rangle\,d\alpha
\bigg]\,d\x,
\end{equation}
and a transport contribution
\begin{equation}
{\q_T(\r,t)}\equiv\Big\langle \sum_{j\neq k}\frac{1}{2}V_{jk}(|\r_j-\r_k|)(\v_k-\v)\delta(\r_k-\r)\Big\rangle
+\Big\langle\sum_{k}V^\mathrm{e}_k(\r_k)(\v_k-\v)\delta(\r_k-\r)\Big\rangle.
\end{equation}
\end{widetext}
%


\textit{Energy balance with collective observables.}
The time derivatives of the collective energy densities $\epsilon_s$ and $\epsilon_{p_s}$ defined in \eqref{eq:eps-til} and \eqref{eq:es-til} are 
\begin{equation}
\frac{d \epsilon_{p_s}}{dt}=\mbox{}   {\Big\langle\frac{p_s^3}{2Q^2\omega_\mathrm{ref}}\Big\rangle
+\Big\langle\frac{p_s}{Q\omega_\mathrm{ref}}\Big[\sum_km_k{|\v_k|^2}-A\Big]\Big\rangle}      \equiv \sigma_{\epsilon}^{p_s}
\end{equation}
and
\begin{equation}
\frac{d \epsilon_s}{dt}=\mbox{}    \Big\langle \frac{p_s}{Q\omega_\mathrm{ref}}A\ln s\Big\rangle    \equiv \sigma_{\epsilon}^{s};
\end{equation}
these derivatives induce two uniform source terms, $\sigma_{\epsilon}^{p_s}$ and $\sigma_{\epsilon}^{s}$.

All in all, the balance of the energy density $\epsilon$ defined in \eqref{eq:energy-density-til} takes the form
\begin{equation}\label{eq:energy-balance-til}
\frac{\de{\epsilon}}{\de t}=-\dvg{\r}(\q+\epsilon_K\v+\epsilon_V\v-\T^\tsp\v)+\sigma_\epsilon,
\end{equation}
%
where $\q$ and $\sigma_\epsilon$ are given by
\begin{align}
\q&\mbox{}=\q_K+\q_V+\q_T,
\\[4pt]
\sigma_\epsilon&\mbox{}=\sigma_\epsilon^0+\sigma_\epsilon^K+\sigma_\epsilon^V+\sigma_\epsilon^{p_s}+\sigma_\epsilon^s.
\end{align}
In addition to the  time-dependent but spatially uniform sources $\sigma_{\epsilon}^{p_s}$ and $\sigma_{\epsilon}^{s}$, $\sigma_\epsilon$ features the classical term
%
\begin{equation}\label{eq:c}
\sigma_\epsilon^0(\r,t)\equiv\Big\langle\sum_k\frac{\de V_k^\mathrm{e}}{\de \r_k}\cdot\v_k\delta(\r_k-\r)\Big\rangle
\end{equation}
and two other new terms, namely
\begin{equation}
\sigma_\epsilon^K(\r,t)\equiv\Big\langle -\frac{p_s}{Q}\sum_k\frac{m_k}{2}|\v_k|^2\delta(\r_k-\r)\Big\rangle
\end{equation}
and
\begin{multline}
\sigma_\epsilon^V(\r,t)\equiv
\Big\langle \frac{p_s}{Q}\sum_{j\neq k}\frac{1}{2}V_{jk}(|\r_j-\r_k)|)\delta(\r_k-\r)\Big\rangle\\
+\Big\langle \frac{p_s}{Q}\sum_{k}V^\mathrm{e}_k(\r_k)\delta(\r_k-\r)\Big\rangle.
\end{multline}
While the time evolution of $\epsilon_K$ and $\epsilon_V$ produces spatially dependent diffusive fluxes, convective terms, and sources, the time evolution of $\epsilon_{p_s}$ and $\epsilon_{s}$ generates only uniform sources.
\vskip 6pt
\textit{Energy balance with distributed observables.}
To deduce the balance of the energy density $\bar\epsilon$ defined in \eqref{eq:energy-balance-tim}, we compute the time derivatives of the distributed observables $\bar{\epsilon}_{p_s}$ and $\bar{\epsilon}_{s}$ defined in \eqref{eq:kps} and \eqref{eq:esnh}. Specifically, we find that
\begin{equation}
\frac{\de \bar{\epsilon}_{p_s}}{\de t}
=-\dvg{\r}(\q_{p_s}+\bar{\epsilon}_{p_s}\v)+\sigma_{\bar{\epsilon}}^{p_s},
\end{equation}
{with}
\begin{equation}
\q_{p_s}(\r,t)\equiv\Big\langle \frac{p_s^2}{2Q}\sum_k\frac{1}{N}(\v_k-\v)\delta(\r_k-\r)\Big\rangle
\end{equation}
and
\begin{multline}
\sigma_{\bar{\epsilon}}^{p_s}(\r,t)\equiv\Big\langle \frac{p^3_s}{2Q^2}\sum_k\frac{1}{N}\delta(\r_k-\r)\Big\rangle\\
+\Big\langle\frac{p_s}{Q}\Big[\sum_j m_j{|\v_j|^2}-A\Big]\sum_k\frac{1}{N}\delta(\r_k-\r)\Big\rangle.
\end{multline}
Additionally, we find that
\begin{equation}
\frac{\de \bar{\epsilon}_s}{\de t}
=-\dvg{\r}(\q_s+\bar{\epsilon}_s\v)+\sigma_{\bar{\epsilon}}^s,
\end{equation}
{with}
\begin{equation}
\q_s(\r,t)\equiv\Big\langle A(\ln s-1)\sum_k\frac{1}{N}(\v_k-\v)\delta(\r_k-\r)\Big\rangle
\end{equation}
and
\begin{equation}
\sigma_{\bar{\epsilon}}^s(\r,t)\equiv\Big\langle\frac{p_s}{Q}A\ln s\sum_k\frac{1}{N}\delta(\r_k-\r)\Big\rangle.
\end{equation}

All in all, the balance of the energy density $\bar\epsilon$ reads
\begin{equation}\label{eq:energy-balance}
\frac{\de\bar{\epsilon}}{\de t}=-\dvg{\r}(\q+\bar{\epsilon}\v-\T^\tsp\v)+\sigma_{\bar{\epsilon}},
\end{equation}
where 
$\q$, and $\sigma_{\bar{\epsilon}}$ are given by
\begin{align}
\q=\mbox{}&\q_K+\q_V+\q_T+\q_{p_s}+\q_s,
\\[4pt]
\sigma_{\bar{\epsilon}}=\mbox{}&\sigma_\epsilon^0+\sigma_\epsilon^K+\sigma_\epsilon^V+\sigma_{\bar{\epsilon}}^{p_s}+\sigma_{\bar{\epsilon}}^s.
\end{align}
In this case, each of the energy densities produce flux and convective terms influencing the energy balance \eqref{eq:energy-balance}.

\subsection{Application to the APR extended Hamiltonian}

In this section, we apply our generalized IKN procedure to an extended Hamiltonian system akin to but more general than that considered by Parrinello \& Rahman,\cite{PR1,*PR2} whose successful predictions of  stress-induced displacive phase transitions in certain crystalline materials initiated an irreversible change in format in the MD simulation of such phenomena. We reiterate that we chose this application not only for its intrinsic importance, but also stimulated by recently proposed multiscale numerical schemes\cite{Ulz15,LiTong15,*TongLi15} for coupling atomistic and continuum models. Importantly, those numerical schemes are based on an APR approach, and we view determining the form of the implied continuum balances to be an endeavor of particular importance and interest. In a departure from previous treatments, we do so by employing the general and exact form of the kinetic energy, a form which reduces to that postulated by Parrinello \& Rahman under specific conditions analyzed by Podio-Guidugli.\cite{ppg1} We regard maintaining this level of generality as important for obtaining a consistent and unprejudiced physical description at different scales.  

In what follows, we denote by a simple juxtaposition the contraction of a single tensorial index between a tensor and a vector or between two tensors. For both tensors and vectors, we denote by $\boldsymbol A\cdot\boldsymbol B$, with a single centered dot, the scalar product defined through the trace operator by $\mathrm{tr}(\boldsymbol A^\tsp\boldsymbol B)$.

\subsubsection{The extended Hamiltonian}

The additional degrees of freedom that extend the ordinary Lagrangian within an APR framework are those used to describe the deformations of the computational cell; they comprise an invertible second-order tensor $\F$, with positive determinant, which maps the referential position vector $\s_k$ of the $k$-th material point into its current position vector $\r_k$:
\begin{equation}\label{eq:APR0}
\r_k=\F\s_k.
\end{equation}
Accordingly, the Lagrangian coordinates of this system are given by $q=((\s_k)_{k=1}^N,\F)$. 
%
%
The time-dependence of $\F$ renders \eqref{eq:APR0} a generalization of the Cauchy--Born rule, from which it follows that
\begin{equation}\label{eq:APRv}
\dot{\r}_k=\F\dot\s_k+\dot\F\s_k=\v_k.
\end{equation}
As we shall see, \eqref{eq:APRv} is crucial for a correct definition of physical observables. Importantly, it must be used to compute the Lagrangian form of the kinetic energy:
\begin{multline}
K_\mathrm{L}:=\frac{1}{2}\sum_k m_k|\F\dot\s_k+\dot\F\s_k|^2
\\
=\frac{1}{2}\F^\tsp\F\cdot\mskip-2mu\sum_km_k\dot\s_k\otimes\dot\s_k+\frac{1}{2}\dot{\F}^\tsp\dot{\F}\cdot\mskip-2mu\sum_k m_k\s_k\otimes\s_k
\\
+\frac{1}{4}(\dot{\F}^\tsp\F+\F^\tsp\dot{\F})\cdot\mskip-2mu\sum_k m_k(\dot\s_k\otimes\s_k+\s_k\otimes\dot\s_k).
\mskip-12mu
\end{multline}
The Lagrangian kinetic energy $K_L$ is a symmetric and positive semi-definite quadratic form in the variables $((\dot\s_k)_{k=1}^N,\dot\F)$, which depends  parametrically  on $q$. The {corresponding} Hamiltonian kinetic energy $K$ is the Legendre--Fenchel transform of $K_\mathrm{L}$. Although $K_\mathrm{L}$ depends on $\s_k$ and $\dot{\s}_k$, it is independent of $\r_k$ and, therefore, so is $K$.
We thus have
\begin{equation}\label{eq:LFprop}
\p_k=\frac{\de K_\mathrm{L}}{\de\dot{\s}_k},\;\G=\frac{\de K_\mathrm{L}}{\de\dot\F},\; 
\dot\s_k=\frac{\de K}{\de \p_k},\text{ and }\dot\F=\frac{\de K}{\de\G}.
\end{equation}
The conjugate momentum $\p_k$ and the physical momentum $m_k\v_k$ of the $k$-th material point are linked by the transformation
\begin{equation}\label{eq:pk}
\p_k=m_k\F^\tsp\dot\r_k=m_k\F^\tsp(\F\dot\s_k+\dot\F\s_k),
\end{equation}
from which we deduce the alternative expression $\F^{-\tsp}\p_k$ for the physical momentum of the $k$-th material point.

The APR extended Hamiltonian that we consider is
\begin{equation}\label{eq:Hapr}
H_\mathrm{APR}(q,p)=K(q,p)+V(q)-\omega_{\mathrm{ref}}\P\cdot\F,
\end{equation}
with $p=((\p_k)_{k=1}^N,\G)$. Here, as usual, the potential energy depends on the pairwise interactions of material points in the positions they currently occupy:
\begin{equation}
V(q):=\sum_{j\neq k}\frac{1}{2}V_{jk}(|\F(\s_j-\s_k)|).
\end{equation}
Given our present purposes, the addition of an external potential would not add anything of importance. What makes an approach of APR type interesting is the last term of \eqref{eq:Hapr},
$-\omega_\mathrm{ref}\P\cdot\F$, which is of an enthalpic character. {In that term,} $\omega_\mathrm{ref}>0$ denotes the constant reference volume of the computational cell, while the time-independent control parameter $\P$ specifies the prescribed macroscopic Piola stress in the referential configuration of the computational domain throughout the MD simulation.

In this extended Hamiltonian system, the evolution of any probability density $\tilde\rho(q,p,t)$ is determined by Liouville's equation \eqref{eq:LiouvilleE}, which takes the following explicit form:
\begin{multline}
\frac{\de\tilde\rho}{\de t}=\sum_k\bigg( \frac{\de H_\mathrm{APR}}{\de\s_k}\cdot\frac{\de\tilde\rho}{\de\p_k}-\frac{\de\tilde\rho}{\de\s_k}\cdot\frac{\de H_\mathrm{APR}}{\de\p_k}\bigg)\\+\frac{\de H_\mathrm{APR}}{\de\F}\cdot\frac{\de\tilde\rho}{\de\G}-\frac{\de\tilde\rho}{\de\F}\cdot\frac{\de H_\mathrm{APR}}{\de\G}.
\end{multline}

\subsubsection{Definition of the observables}

The discussion in the previous sections provides us with precise indications about the appropriate form for the various observables. We define the mass density
\begin{equation}
\rho(\r,t)\equiv\Big\langle\sum_k m_k\delta(\F\s_k-\r)\Big\rangle,
\end{equation}
the linear momentum density
\begin{align}
\rho\v(\r,t)&\mbox{}\equiv\Big\langle\sum_k \F^{-\tsp}\p_k\delta(\F\s_k-\r)\Big\rangle\notag\\
&\mbox{}=\Big\langle\sum_k m_k\v_k\delta(\F\s_k-\r)\Big\rangle,
\end{align}
the kinetic energy density
\begin{align}
\epsilon_K(\r,t)&\mbox{}\equiv\Big\langle\sum_k \frac{1}{2m_k}|\F^{-\tsp}\p_k|^2\delta(\F\s_k-\r)\Big\rangle\notag\\
&\mbox{}=\Big\langle\sum_k \frac{m_k}{2}|\v_k|^2\delta(\F\s_k-\r)\Big\rangle,\label{eq:eKapr}
\end{align}
and the potential energy density
\begin{equation}\label{eq:eVapr}
\epsilon_V(\r,t)\equiv\Big\langle \sum_{j\neq k}\frac{1}{2}V_{jk}(|\F(\s_j-\s_k)|)\delta(\F\s_k-\r)\Big\rangle.
\end{equation}
In keeping with the treatment of Nos\'e's extended Hamiltonian \eqref{eq:Hnh}, we can further define the collective observable
\begin{equation}
\epsilon_{\P}(\r,t)\equiv
-\langle\P\cdot\F\rangle
=-\P\cdot\langle\F\rangle,
\label{eq:enth-til}
\end{equation}
and its distributed counterpart
\begin{equation}
\bar{\epsilon}_{\P}(\r,t)\equiv
-\Big\langle \omega_\mathrm{ref}(\P\cdot\F)\sum_{k}\frac{1}{N}\delta(\F\s_k-\r)\Big\rangle\label{eq:enth}.
\end{equation}

\subsubsection{The continuum balances}

In the case of the APR extended Hamiltonian, the classical form of the mass balance is preserved, as is the form of the linear momentum balance, in which the stress tensor $\T=\T_K+\T_V$ is obtained from \eqref{eq:TK}--\eqref{eq:TVJ} modulo the substitutions
\begin{align}
\r_k&\mapsto\F\s_k,\label{eq:subs1}
\\[4pt]
\frac{\p_k}{s}&\mapsto m_k(\F\dot\s_k+\dot\F\s_k)=\F^{-\tsp}\p_k=m_k\v_k.\label{eq:subs2}
\end{align}

Choosing the collective form \eqref{eq:enth-til} for the enthalpic energy density generates the energy balance
\begin{equation}\label{eq:energy-balance-APR-til}
\frac{\de \epsilon}{\de t}=-\dvg{\r}(\q+\epsilon_K\v+\epsilon_V\v-\T^\tsp\v)+\sigma_\epsilon,
\end{equation}
with
\begin{align}
\epsilon&=\epsilon_K+\epsilon_V+\epsilon_{\P},
\\[4pt]
\q&=\q_K+\q_V+\q_T,
\end{align}
and
\begin{equation}
\sigma_\epsilon\equiv\dot\epsilon_P.
\end{equation}
In these relations, the terms with suffixes $K$, $V$, and $T$ have the form of the corresponding terms in the application of the IKN procedure to the NH extended Hamiltonian. If the computational cell is viewed as a homogeneously deforming elastic body, then $\omega_{\mathrm{ref}}\F$ constitutes an extensive variable and $\sigma_\epsilon$ provides a reckoning of the power expenditure associated with changing the shape of the computational cell under the influence of the Piola stress $\P$.\cite{McLellan,Man}

If we choose instead the distributed form \eqref{eq:enth} for the enthalpic energy density, the energy balance reads
\begin{equation}\label{eq:energy-balance-APR}
\frac{\de \bar{\epsilon}}{\de t}=-\dvg{\r}(\q+\bar{\epsilon}\v-\T^\tsp\v),
\end{equation}
where
\begin{equation}
\bar{\epsilon}=\epsilon_K+\epsilon_V+\bar{\epsilon}_{\P},
\end{equation}
and
\begin{equation}
\q=\q_K+\q_V+\q_T+\q_{\P}.
\end{equation}
The stress-related terms $\epsilon_{\P}\v$ and $\q_{\P}$ are both
due to the enthalpic term in the extended Hamiltonian. {The former has a convective nature; the latter, whose form is}
\begin{equation}
\q_{\P}\equiv-\Big\langle\omega_\mathrm{ref}(\P\cdot\F)\sum_k\frac{\v_k-\v}{N}\delta(\F\s_k-\r)\Big\rangle,
\end{equation}
accounts for the heat flux generated by the action of the prescribed macroscopic Piola stress} $\P$.

\subsubsection{About the use of the Parrinello--Rahman kinetic energy}

The Lagrangian kinetic energy introduced by Parrinello \& Rahman\cite{PR1,*PR2} is
\begin{equation}\label{eq:PRke}
K_\mathrm{PR}:=
\frac{1}{2}\,\F^\tsp\F\cdot\mskip-2mu\sum_km_k\dot\s_k\otimes\dot\s_k+ \frac{1}{2}W|\dot{\F}|^2,\quad W>0.
\end{equation}
Since $K_\mathrm{PR}$ is a symmetric and positive definite quadratic form in the variables $((\dot\s_k)_{k=1}^N,\dot\F)$, the above analysis invoving $K_\mathrm{L}$ is applicable also in this case, with similar results. What changes is not the form of the resulting continuum balances but the physical significance of the terms in those balances that depend directly on the microscopic evolution equations derived from the choice of $K_\mathrm{PR}$ for the kinetic energy. 

It is worth observing that the inertial parameter $W$ plays a role akin to that of $Q$ in the Nos\'e--Hoover Hamiltonian. This marks an important distinction between the extension pertinent to the Nos\'e--Hoover Hamiltonian, which would be paralleled by one based on the Parrinello--Rahman Lagrangian, and the extension realized in the APR Hamiltonian discussed in this paper. When the physical kinetic energy is translated in terms of the extended variables, the outcome is $K_{L}$ and there is no need for inertial parameters, whose value is regarded as adjustable in more than one way.\cite{HCA,PR2} In this connection, we also observe that the difference between $K_\mathrm{L}$ and $K_\mathrm{PR}$ vanishes if two constraints are imposed on the  motion of the computational cell and the particles therein,\cite{ppg1} one being holonomic and of the form
\begin{equation}
\sum_k m_k\s_k\otimes\s_k=W\I,
\end{equation}
where $\I$ is the identity tensor, and the other being non-holonomic and of the form
\begin{equation}
\dot{\F}\F^{-1}=(\dot{\F}\F^{-1})^\tsp.
\end{equation}
It would be of some interest to introduce an extended Hamiltonian in which \emph{ad hoc} Lagrangian multipliers would mediate the presence of these constraints.

\section{Conclusions}\label{sec:conclusions}

We discussed the application of the Irving--Kirkwood--Noll procedure to extended Hamiltonian systems employed in certain molecular dynamics simulations. This procedure makes it possible to obtain macroscopic balance laws of continuum mechanics that can be regarded as specific collective counterparts of the extended Hamiltonian dynamics of a system of material points.

Our analysis demonstrates that the presence of auxiliary degrees of freedom, needed to control the statistical properties at the atomistic level, affects the structure of the balance equations at the continuum level. Precisely, the modifications induced at the microscopic scale influence the thermodynamic and transport properties at the macroscopic scale by way of novel source and flux terms in the continuum balances for the mass and energy densities.
 
We considered in detail two important examples of extended Hamiltonian systems. First, we discussed the application of the Irving--Kirkwood--Noll procedure to the extended Hamiltonian underlying the Nos\'e--Hoover thermostat. Notably, the effective thermal inertia present in that model produces source terms in the balances of both mass and energy, source terms which do not feature in standard continuum models for isothermal processes. Second, we discussed the application of the Irving--Kirkwood--Noll procedure to an extended Hamiltonian that we designated after Andersen, Parrinello, and Rahman because they were first to introduce extended Lagrangians to control the pressure and stress imposed on a collection of material points during a molecular dynamics simulation. In this context, it is possible to identify a contribution to the heat flux that is absent in the classical context. The term in question encodes the effects on the macroscopic balance of the energy needed to keep a constant the stress applied to the system. 

Our findings are particularly relevant when considering computational schemes aiming at connecting the atomistic and continuum scales. Indeed, we view it as essential to correctly subsume the statistical properties imposed on the microscopic model on the thermodynamic and transport properties of the coupled continuum. When developing multiscale numerical schemes that aim at coupling, in a physically consistent manner, statistical information retrieved from MD with continuum modeling, this goal can be achieved by adopting continuum mechanical balances of the type we derived. We surmise that a failure to do so may lead to undesirable inconsistencies and computational artifacts.

\section*{Acknowledgments}
The authors wish to acknowledge the support of CECAM -- Centre Europ\'een de Calcul Atomique et Mol\'eculaire through the organization of a workshop that fostered the development of the present study. E.\ F.\ and G.\ G.\ G.\ gratefully acknowledge support from the Okinawa Institute of Science and Technology Graduate University with subsidy funding from the Cabinet Office, Government of Japan.


%

\end{document}